\documentclass[aps, prd, superscriptaddress, twocolumn,a4paper]{revtex4-1}

\usepackage{graphicx}
\usepackage{amsmath}
\usepackage{color}
\usepackage{float}
\usepackage{bm}
\usepackage[english]{babel}

\usepackage{ulem}
\usepackage{appendix}
\usepackage[colorlinks=true,linkcolor= red, citecolor = cyan, urlcolor = teal ]{hyperref}
\usepackage{epstopdf}

\usepackage{subfig}
\usepackage{comment}

\usepackage[dvipsnames]{xcolor}
\newcommand{\be}{\begin{equation}}
\newcommand{\ee}{\end{equation}}
\newcommand{\ba}{\begin{eqnarray}}
\newcommand{\ea}{\end{eqnarray}}
\newcommand{\nn}{\nonumber\\}

\begin{document}

\title{Impact of Medium Anisotropy on Quarkonium Dissociation and Regeneration}

\author{Captain R. Singh}
\email{captainriturajsingh@gmail.com}
\affiliation{Department of Physics, Indian Institute of Technology Indore, Simrol, Indore 453552, India}

\author{Mohammad Yousuf Jamal}
\email{Corresponding Author: yousufjml5@gmail.com}
\affiliation{School of Physical Sciences, Indian Institute of Technology Goa, Ponda 403401, Goa, India}

\author{Raghunath Sahoo}
\email{Corresponding Author: Raghunath.Sahoo@cern.ch}
\affiliation{Department of Physics, Indian Institute of Technology Indore, Simrol, Indore 453552, India}

\begin{abstract}

Quarkonium production in ultra-relativistic collisions plays a crucial role in probing the existence of hot QCD matter. This study 
explores quarkonia states dissociation and regeneration in the hot QCD medium while considering momentum anisotropy. The net quarkonia 
decay width ($\Gamma_{D}$) arises from two essential processes: collisional damping and gluonic dissociation. The quarkonia regeneration 
includes the transition from octet to singlet states within the anisotropic medium. Our study utilizes a medium-modified potential that 
incorporates anisotropy via particle distribution functions. This modified potential gives rise to collisional damping for quarkonia due 
to the surrounding medium, as well as the transition of quarkonia from singlet to octet states due to interactions with gluons. 
Furthermore, we employ the detailed balance approach to investigate the regeneration of quarkonia within this medium. Our comprehensive 
analysis spans various temperature settings, transverse momentum values, and anisotropic strengths. Notably, we find that, in addition to 
medium temperatures and heavy quark transverse momentum, anisotropy significantly influences the dissociation and regeneration of various 
quarkonia states.

\end{abstract}
 \date{\today}
\maketitle

\section{Introduction}
\label{intro}
 
 The relativistic heavy-ion collision (HIC) experiments at the Relativistic Heavy Ion Collider (RHIC) and the Large Hadron Collider (LHC) 
 offered an opportunity to explore the properties of hot dense QCD matter known as the Quark Gluon Plasma (QGP)~\cite{expt_rhic1,expt_rhic2,expt_rhic3,expt_rhic4,expt_lhc1,expt_lhc2,expt_lhc3}. The transient nature of 
 this phase, resulting from the confinement properties of Quantum Chromodynamics (QCD), necessitates the study of discernible signatures 
 and probes. One of the most prominent and analyzed signatures is the suppression of heavy quarkonia ($q-\bar{q}$) due to their 
 interactions with the QGP medium~\cite{Matsui:1986dk, McLerran:1986zb}. The pioneering work on the 
 dissolution of various quarkonia states, attributed to color screening within the finite-temperature QGP medium, was first undertaken by 
 Matsui and Satz~\cite{Matsui:1986dk}. Since then, a substantial body of literature has contributed critical refinements to this 
 research~\cite{pks1, Mocsy:2004bv, Agotiya:2008ie, Patra:2009qy, Datta:2003ww, Brambilla:2008cx, Burnier:2009yu, Dumitru:2009fy, 
 Laine:2006ns, crs1}. Interestingly, an expected suppression of $J/\psi$ was observed at the Super Proton Synchrotron (SPS) at CERN 
 \cite{NA38:1989yxm}. However, this suppression did not exhibit a corresponding increase with higher beam energies at RHIC 
 \cite{STAR:2009irl}. To address this puzzle, the concept of recombination of $c-\bar{c}$ pairs into $J/\psi$ within the QGP medium was 
 introduced. Consequently, recombination processes play an important role in understanding the observed number of $J/\psi$ particles in 
 experiments~ \cite{Silvestre:2008tw, Scardina:2013nua}.\\

In HIC, the quarkonia, characterized by their considerable mass, predominantly emerge during the initial stages directly following the 
collisions. The production and evolution of quarkonia can be categorized into two distinct processes: (i) perturbative production: the 
generation of quark and anti-quark ($q {\bar q}$) pair through nucleon-nucleon collisions constitutes the perturbative aspect of 
quarkonium production~\cite{bodwin1}. (ii) Non-perturbative evolution: the subsequent formation and development of bound-state from these 
$q {\bar q}$ pairs are governed by non-perturbative QCD phenomena. This dual nature of quarkonia allows us to explore both perturbative 
and non-perturbative QCD effects. Theoretical frameworks grounded in QCD, such as  non-relativistic QCD (NRQCD), potential NRQCD/
(pNRQCD)~\cite{bodwin2,bodwin3, Brambilla:1999xf, Brambilla:2004jw} and fragmentation approaches~\cite{kang1,kang2}, are widely employed to investigate quarkonium production.\\

It is reported that the stability of quarkonium states can be compromised by various QGP effects such as color 
screening ~\cite{Matsui:1986dk, crs1}, collisional damping, and gluonic dissociation~\cite{nendzig, wols, 
sharma,crs2}, {\it etc}. The static potential between $q-\bar{q}$ pair placed in a QCD medium consists of two 
parts; the first part represents the real part of the potential, and the other part of the potential is the 
imaginary part. The collisional damping arises due to the imaginary part of the potential between $q-{\bar q}$ 
pairs in the time limit,  $t\rightarrow\infty$. It represents the thermal decay width induced by collisional 
(Landau) damping of the low-frequency gauge fields that mediate interactions between heavy $q-\bar{q}$ 
pair.~\cite{Laine:2006ns}. The gluonic dissociation occurs when quarkonia absorb soft gluons ($E1$  gluons), leading to a transition from a 
color singlet state to a color octet state—an unbound quark-antiquark pair~\cite{Brambilla:2008cx, 
peskin, bhanot}. The mechanism governing quarkonia production in the color singlet state has been explored in Ref.~\cite{Adamczyk:2012ey, 
Berger:1980ni, Rapp:2008tf}. Furthermore, the regeneration of quarkonia within the QGP medium is also possible and 
offers two primary 
avenues: (i) Uncorrelated $q-{\bar q}$ recombination: Free flowing $q$ and $\bar{q}$ produced within the medium, through 
various different processes, 
have the potential to merge into open mesons or hidden quarkonia through the recombination of $q$ and $\bar{q}$. The 
production of quarkonia through the recombination of $q$ and $\bar{q}$ is considered as regeneration of quarkonia 
through uncorrelated recombination of quark and anti-quark~\cite{pbm, andronic,rapp1,rapp2,thews1,thews2,thews3}. It is 
particularly significant for charmonium states like $J/\psi, \chi_{c}, \psi^{'}$, {\it etc.,} at LHC energies due to the 
relatively higher production of $c$ and $\bar{c}$ in HICs compared to $b$ and $\bar{b}$ quarks. One may use the 
coalescence mechanism to calculate the formation of quarkonia from such quarks and anti-quarks. (ii) Correlated 
$q-{\bar q}$ recombination: this process reverses the gluonic dissociation, which produces a color octet state. The 
color octet state represents an unbound configuration of $q-\bar{q}$ pairs, and an octet potential is introduced to account for such configurations~\cite{bram}, establishing the spatial correlation between $q-\bar{q}$ pairs. The transition of these pairs from a color octet to a color singlet state is regarded as quarkonia regeneration, attributed to the correlated nature of the $q-\bar{q}$ pair. In this particular situation, it is worth mentioning that besides the charmonia, there is also the possibility for the regeneration of bottomonia via the de-excitation of correlated $b-{\bar b}$ pairs~\cite{crs2,crs4}. It is to be noted that the regeneration of quarkonia may happen at any point in time, given that the medium temperature should be less than the dissociation temperature of quarkonia.\\

Following the UMQS model~\cite{crs1,crs2,crs3,crs4,crs5}, centered on the formulation of gluonic dissociation and 
collisional damping within an isotropic medium, the current study extends its scope to accommodate momentum 
anisotropy. Several researchers have delved into the dissociation of quarkonia, exploring diverse facets, including 
momentum anisotropy. However, the prospect of recombination in the context of an anisotropic QGP medium remains largely 
unexplored. This aspect is important because a significant increase in the production of $c\bar{c}$ and higher $R_{AA}$ 
for $J/\psi$ at low momentum in experiments has been noticed at LHC. These findings strongly suggest that quarkonium, 
like $J/\psi$, might be created through recombination either at the boundary between phases or during the QGP phase 
\cite{Andronic:2017pug}. Although the decay width has been investigated previously by some of us \cite{Jamal:2018mog}, our study represents the inaugural attempt to devise a methodology for investigating this recombination process while considering momentum anisotropy. This anisotropy remains prevalent throughout all stages, particularly in the context of non-central collisions. Consequently, our objective is to evaluate the decay width and regeneration for $J/\psi$ and $\Upsilon$(1S) within the QGP medium, considering the presence of momentum anisotropy. The exploration of momentum anisotropy has been a focal point in various contexts, as evidenced by studies such as \cite{Romatschke:2006bb, Schenke:2006yp, Mauricio:2007vz, Dumitru:2007rp, Baier:2008js, Dumitru:2007hy, Carrington:2008sp}.
We adopt an approach inspired by Refs.~\cite{Romatschke:2003ms, Carrington:2014bla, Jamal:2017dqs}, where anisotropy is introduced at the level of the distribution function by stretching and squeezing it along one direction. Subsequently, we derive the medium-modified Cornell potential, considering 
anisotropic dielectric permittivity. This modification yields the potential as a complex quantity consisting of finite real and imaginary parts~\cite{Laine:2006ns, Beraudo:2007ky, Laine:2007qy, hatsuda, Margotta:2011ta, Strickland:2011aa, Thakur:2013nia}. Furthermore, along with the anisotropic effects, we have also observed the influence of relativistic Doppler shift on both, {\it i.e.,} the particle's net decay width ($\Gamma_{\rm D}$) and recombination reactivity ($\Gamma_{\rm F}$) which is responsible for quarkonia regeneration. Transverse momentum ($p_{\rm T}$) and temperature (T) dependent $\Gamma_{\rm D}$, $\Gamma_{\rm F}$ are obtained for $J/\psi$, $\Upsilon$(1S) for anisotropic strength, $\zeta =  0.2, 0.4$ and T  = 200, 300, 400 MeV. This study provides a comprehensive method to estimate the net quarkonia suppression. It can be extended to obtain the survival probability of quarkonia in ultra-relativistic collisions at the RHIC and LHC energies. However, it is beyond the scope of the current analysis; we leave it for future investigations. \\

The manuscript is structured as follows: Section~\ref{HQP} provides the formalism for the dissociation and correlated recombination of various quarkonia states in the presence of momentum anisotropy. Section~\ref{RaD} is dedicated to presenting our results and engaging in discussions. Finally, in Section~\ref{SaF}, we summarize our findings and contemplate future aspects of the current work. Throughout the text, we employ natural units with $c=k_B=\hbar=1$. Three vectors are in bold typeface, while four are in regular font. The center dot denotes the four-vector scalar product, with the metric tensor specified as $g_{\mu\nu}={\text {diag}}(1,-1,-1,-1)$.

\section{Formalism}
\label{HQP}

The interaction between a ($q-{\bar q}$) pair is characterized by the vacuum Cornell potential, which includes both a Coulombic and a string component~\cite{Eichten:1978tg, Eichten:1979ms}. This potential is expressed as:

\ba
{\text V(r)} = -\frac{\alpha_s}{r}+\sigma r.
\ea

Here, $r$ represents the radius of the quarkonia state under investigation, $\alpha_s$ denotes the strong coupling constant, and $\sigma$ represents the string tension. To account for medium effects, we modify this potential using the dielectric permittivity, $\epsilon(k)$, which considers the presence of momentum anisotropy. In Fourier space, this modification is formulated as:

\ba
\grave{V}(k)=\frac{{\bar{\text V}}(k)}{\epsilon(k)},
\label{eq:v}
\ea

where ${\bar{\text V}}(k)$ is the Fourier transform of ${\text V(r)}$ and is obtained as:

\ba
{\bar{\text V}}(k)= -\sqrt\frac{2}{\pi}\bigg(\frac{\alpha_s}{k^2}+2\frac{ \sigma}{k^4}\bigg).
\ea

Here, the transition from the hadronic phase to the QGP is assumed to be a crossover~\cite{Rothkopf:2011db}, ensuring that the string tension does not abruptly disappear at or near the critical temperature, $T_c$. As mentioned earlier, some of us have derived the medium-modified potential in the presence of anisotropy in Ref.~\cite{Jamal:2018mog}; we briefly provide the necessary steps for completeness in the upcoming sections.

\subsection{Dielectric Permittivity in the Anisotropic QGP Medium}

The anisotropy in this formalism is introduced through the medium's particle distribution functions, represented as $f_{g}(p)$ and $f_{q/{\bar q}}(p)$:

\ba
f_{g}(p) = \frac{1}{e^{\beta E_g} - 1}, ~~~ f_{q/{\bar q}}(p) = \frac{1}{e^{\beta E_p} + 1}.
\ea

Here, $\beta = 1/T$ ($T$ is the temperature of the system in the units of energy). For the gluons, $E_{g}=|{\bf p}|$  and $E_{q} = \sqrt{|{\bf p}|^2+m_q^2}$ for the quarks/ antiquarks and $m_q$ denotes their mass. ${\bf p}$ represents the medium quarks and gluons momentum (not to mix up with heavy quark momentum later). We adopt the method utilized in these Refs.~\cite{Romatschke:2003ms, Carrington:2014bla, Jamal:2017dqs}, where anisotropic distribution functions are derived from isotropic ones through rescaling in one direction in momentum space:

\ba
f({\mathbf{p}})\rightarrow f_{\zeta}({\mathbf{p}}) = C_{\zeta}~f(\sqrt{{\bf p}^{2} + \zeta({\bf p}\cdot{\bf \hat{n}})^{2}}).
\ea

Here, ${\bf{\hat{n}}}$ is a unit vector (${\bf \hat{n}}^{2} = 1$) representing the direction of anisotropy, and $\zeta$ quantifies the anisotropic strength in the medium. It describes the degree of squeezing ($\zeta > 0$, oblate form) or stretching ($-1<\zeta<0$, prolate form) along the ${\bf \hat{n}}$ direction. Normalizing the Debye mass $m_D$, defined as  \cite{Carrington:2014bla}:

\ba
m^{2}_D &=& 4\pi \alpha_s \bigg(-2N_c \int \frac{d^3 p}{(2\pi)^3} \partial_p f_g(p)\nn
&-& N_f \int \frac{d^3 p}{(2\pi)^3} \partial_p \left(f_q(p)+f_{\bar q}(p)\right)\bigg).
\ea

In both isotropic and anisotropic media, yields the normalization constant $C_{\zeta}$:

\ba
  C_{\zeta}= 
\begin{cases}
   \frac{\sqrt{|\zeta|}}{\tanh^{-1}\sqrt{|\zeta|}}& \text{if }~~ -1\leq\zeta<0\\
    \frac{\sqrt{\zeta}}{\tan^{-1}\sqrt{\zeta}}    & \text{if }~~ \zeta\geq0 .
\end{cases}
\ea

In the limit of small $\zeta$, this expression approximates to:

\ba
  C_{\zeta}= 
\begin{cases}
   1-\frac{\zeta}{3} +O\left(\zeta ^{\frac{3}{2}}\right)& \text{if }~~ -1\leq\zeta<0\\
    1+\frac{\zeta}{3} +O\left(\zeta ^{\frac{3}{2}}\right)   & \text{if }~~ \zeta\geq0 .
\end{cases}
\ea

The inclusion of $C_\zeta$ serves the purpose of averting the sudden loss of anisotropy as the momentum of the medium 
particles aligns perpendicularly to the anisotropic direction, particularly as $\theta_n$ approaches $\pi/2$. This 
scenario compromises the physical integrity of the situation. On the contrary, maintaining anisotropy is assured through $C_{\zeta}$ as long as $\zeta\ne 0$. In the context of the current situation, where $\zeta$ is small, the value of $C_\zeta$ approximates 
1.1. This adjustment subtly enhances the fidelity of the physical representation in the model. It was initially set to unity in Ref.\cite{Romatschke:2003ms}. However, in a subsequent study by Romatschke and Strickland \cite{Romatschke:2004au}, it was normalized based on the anisotropic number density to the isotropic one, yielding \(C_\zeta = \sqrt{1+\zeta}\) an alternative expression for the same. 

Before calculating the dielectric permittivity for the anisotropic QGP medium, it is essential to know that perturbative theory at $T>0$ faces infrared singularities and gauge-dependent results due to the incomplete nature of perturbative expansion at these temperatures. However, HTL resummation~\cite{Braaten:1989mz}, semi-classical transport theory, or many-particle kinetic theory yield consistent results up to one-loop order. These methods lead to the same expression for the gluon self-energy, $\Pi^{\mu\nu}$, which in turn affects the medium's dielectric permittivity \cite{Kumar:2017bja}. 
In the static limit, the dielectric permittivity, $\epsilon^{-1}({\bf k})$ can be derived from the temporal component of the gluon propagator considering the Coulomb gauge within the linear response theory as~\cite{Agotiya:2016bqr, Jamal:2018mog}:

\ba
\epsilon^{-1}({\bf k}) = -\lim_{\omega \to 0}k^2\Delta^{00}(\omega,{\bf k}).
\label{eq:eps}
\ea

The temporal components of the real and imaginary parts of the gluon propagator are given as follows:
 
\ba
\Re[\Delta^{00}({\omega = 0,\bf k})] &=& \frac{-1}{k^2+m_{D}^2}-\zeta\Big(\frac{1}{3(k^2+m_{D}^2)}\nn &-&\frac{m_{D}^2(3\cos{2\theta_n}-1)}{6
(k^2+m_{D}^2)^2}\Big),
\label{eq:Re_delta}
\ea

\ba
\Im[\Delta^{00}({\omega = 0,\bf k})] &=& \pi~ T~ m_{D}^2\bigg(\frac{-1}{k(k^2+m_{D}^2)^2} \nn &+& \zeta\Big(\frac{-1}{3k(k^2+m_{D}^2)^2}+\frac{3\sin^{2}{\theta_n}}{4k(k^2+m_{D}^2)^2} \nn &-& \frac{2m_{D}^2\big(3\sin^{2}({\theta_n})-1\big)}{3k(k^2+m_{D}^2)^3}\Big)\bigg).
\label{eq:Im_delta}
\ea

These expressions contain direction and strength to account for the anisotropy in the medium. Now, using Eq.\eqref{eq:Re_delta} and Eq.\eqref{eq:Im_delta} in Eq.\eqref{eq:eps}, respectively the real and imaginary parts of $\epsilon({\bf k})$ can be obtained as,
\ba
\Re[\epsilon^{-1}({\bf k})]&=&\frac{k^2}{k^2+m_{D}^2}+k^2\zeta\Big(\frac{1}{3(k^2+m_{D}^2)}\nn &-&  \frac{m_{D}^2(3\cos{2\theta_n}-1)}{6(k^2+m_{D}^2)^2}\Big),
\label{eq:Re_eps}
\ea
and 
\ba
\Im[\epsilon^{-1}({\bf k})]& =& \pi T  m_{D}^2\bigg(\frac{k^2}{k(k^2+m_{D}^2)^2}-\zeta k^2\Big(\frac{-1}{3k(k^2+m_{D}^2)^2}\nn
&+&\frac{3\sin^{2}{\theta_n}}{4k(k^2+m_{D}^2)^2}-\frac{2m_{D}^2\big(3\sin^{2}({\theta_n})-1\big)}{3k(k^2+m_{D}^2)^3}\Big)\bigg).\nn
\label{eq:Im_eps}
\ea
It is to note that at $\zeta\rightarrow 0$, and  $T\rightarrow 0$, the real part $\Re[\epsilon^{-1}({\bf k})]$ goes to unity, and the imaginary part $\Im[\epsilon^{-1}({\bf k})]$ vanishes, and we get back the vacuum Cornell potential.

 \subsection{Modified Cornell potential in the anisotropic medium}
 
The medium-modified Cornell potential can be obtained using $\epsilon(k)$ in Eq. \eqref{eq:v}. The updated potential in the coordinate space can be found by doing an inverse Fourier transform and is given as,
 \ba 
V(r)= \int \frac{d^3\mathbf{k}}{(2\pi)^{3/2}}(e^{i\mathbf{k} \cdot \mathbf{r}}-1)\grave{V}(k).
 \label{eq:V}
 \ea
 
 Since the medium permittivity is a complex quantity, the updated potential also contains the real and imaginary parts. Employing Eq. \eqref{eq:Re_eps} in Eq. \eqref{eq:v}, the real part of the potential is obtained as,
  \ba
  \Re[V] &=&\alpha_s m_D \left(-\frac{e^{-\grave{r}}}{\grave{r}}-1\right)+\frac{\sigma}{ m_D}  \Big(\frac{2 e^{-\grave{r}}}{\grave{r}}-\frac{2}{\grave{r}}+2\Big)\nn
  &+&\alpha_s m_D \zeta\bigg[-\frac{3 \cos (2 \theta_r )}{2 \grave{r}^3}-\frac{1}{2 \grave{r}^3}+\frac{1}{6}+e^{-\grave{r}} \Big\{\frac{1}{2 \grave{r}^3}\nn
  &+&\frac{1}{2 \grave{r}^2}+\Big(\frac{3}{2 \grave{r}^3}
  +\frac{3}{2 \grave{r}^2}+\frac{3}{4 \grave{r}}+\frac{1}{4}\Big) \cos (2 \theta_r )+\frac{1}{4 \grave{r}}\nn
  &-&\frac{1}{12}\Big\}\bigg]+\frac{\zeta ~ \sigma}{m_D}  \bigg[\left(\frac{6}{\grave{r}^3}-\frac{1}{2 \grave{r}}\right) \cos (2 \theta_r )+\frac{2}{\grave{r}^3}\nn
  &-&\frac{5}{6 \grave{r}}+\frac{1}{3}+e^{-\grave{r}} \Big\{-\frac{2}{\grave{r}^3}-\frac{2}{\grave{r}^2}+\Big(-\frac{6}{\grave{r}^3}-\frac{6}{\grave{r}^2}\nn
  &-&\frac{5}{2 \grave{r}}-\frac{1}{2}\Big) \cos (2 \theta_r )-\frac{1}{6 \grave{r}}+\frac{1}{6}\Big\}\bigg],
  \label{eq:RV}
\ea

where $\grave{r} = r m_D$. Assuming the limit, $\grave{r}\ll1$, Eq.\eqref{eq:RV} becomes,
\ba
\Re[V] &=&\frac{ \grave{r}~ \sigma }{m_D}\left(1+\frac{\zeta }{3}\right)-\frac{\alpha_s~  m_D}{\grave{r}} \bigg(1+\frac{\grave{r}^2}{2}\nn 
&+&\zeta  \left(\frac{1}{3}+\frac{\grave{r}^2}{16}\left(\frac{1}{3}+ \cos \left(2 \theta _r\right)\right)\right)\bigg),
\label{eq:realv}
\ea

here, $\sigma = 0.192$  GeV$^{2}$. The coupling constant is considered as follows; 
\begin{equation}
     \alpha_{s}(\Lambda)  = \frac{6\pi}{(11N_c - 2N_f) \ln\left(\frac{\Lambda}{\Lambda_{MS}}\right)},
     \label{alpha}
\end{equation}
with  $\Lambda_{MS}$ = 0.176 GeV, $N_f = 3$ and $N_c = 3$. Here, the renormalization scale $\Lambda$ under HTL limit is considered $2 \pi T$~\cite{Haque:2014rua}. Now, using Eq.\eqref{eq:Im_eps}, the imaginary part of the potential is found to be:
 {\small \ba
 \Im[V]&=&\frac{\alpha_s ~T ~m^2_D}{2~\pi}\int d^3{\bf k}(e^{i\mathbf{k} \cdot \mathbf{r}}-1)
 \frac{1}{k}\bigg[\frac{-1}{(k^2+m_D^2)^2}\nn
 &+&\zeta[\frac{-1}{3(k^2+m_D^2)^2}+\frac{3\sin^2\theta_n}{2(k^2+m_D^2)^2}
 -\frac{4 m_D^2(\sin^2\theta_n-\frac{1}{3})}{(k^2+m_D^2)^3}]\bigg]\nn
&+&\frac{\sigma T m^2_D}{\pi}\int d^3\mathbf{k}(e^{i\mathbf{k} \cdot \mathbf{r}}-1)
\frac{1}{k^3}\bigg[\frac{-1}{(k^2+m_D^2)^2}\nn
&+&\zeta[\frac{-1}{3(k^2+m_D^2)^2}+\frac{3\sin^2\theta_n}{2(k^2+m_D^2)^2}
-\frac{4 m_D^2(\sin^2\theta_n-\frac{1}{3})}{(k^2+m_D^2)^3}]\bigg].\nn
 \label{eq:im_v2}
 \ea}
 The analytical solutions of Eq.\eqref{eq:im_v2} can be obtained in the limit  $\grave{r}\ll 1$ as,
\ba
\Im[V]&=&\frac{\alpha_s ~ \grave{r}^2~ T}{3} \Big\{\frac{ \zeta }{60} (7-9 \cos 2 \theta_r)-1\Big\}\log \left(\frac{1}{\grave{r}}\right)\nn
&+&\frac{\grave{r}^4 ~\sigma ~ T}{m_D^2}\Big\{\frac{\zeta}{35}  \left(\frac{1}{9}-\frac{1}{4} \cos 2 \theta_r \right)\nn
&-&\frac{1}{30}\Big\}\log \left(\frac{1}{\grave{r}}\right).
 \label{eq:Iim_v3}
 \ea

It is noteworthy that obtaining an analytical solution for the complete imaginary potential described in Eq. \eqref{eq:im_v2} is unattainable. Nevertheless, one can numerically solve it under certain reasonable assumptions. 
 

\subsection{Collisional damping}
\label{CD}

Collisional damping is an inherent characteristic of the complex potential, and as mentioned above, quarkonia potential becomes complex within the medium; the imaginary part of the potential gradually reduces the force fields that bind $q-\bar{q}$ together. Consequently, quarkonia gets dissociated within the QGP medium, and the decay width ($\Gamma_{d,nl}$) corresponding to collisional damping can be obtained by performing a first-order perturbation calculation. The calculation of $\Gamma_{d,nl}$ involves folding the imaginary part of the potential with the radial wavefunction~\cite{crs1}:

\ba
\Gamma_{d,nl}(\tau,p_{T}) = \int g_{nl}(r)^{\dagger}  \Im[V] g_{nl}(r) \, dr.
\label{cold}
\ea

Here, $g_{nl}(r)$ represents the singlet wavefunction corresponding to the specific quarkonia state being investigated, where $n$ and $l$ are the quantum numbers with their usual meanings. For simplicity, some of the authors have chosen the Hydrogen atom wavefunctions \cite{Agotiya:2016bqr, Jamal:2018mog, Nilima:2024nvd}. However, we have obtained these wavefunctions by solving the Schrödinger equation for various $q-\bar{q}$ states.  As the potential is anisotropic, the particle's decay width also contains the respective effect.

\subsection{Gluonic Dissociation}
\label{GD}

The gluon-induced excitation from color singlet to color octet state is defined as ``gluonic dissociation." The 
cross-section corresponding to the process is obtained as given by~\cite{crs3}:

\begin{multline}
\sigma_{gd,nl}(E_g) = \frac{\pi^2\alpha_s^u E_g}{N_c^2}\sqrt{\frac{{m_{Q}}}{E_g
+ E_{nl}}}\\
\;\;\;\;\;\times \left(\frac{l|J_{nl}^{q,l-1}|^2 +
(l+1)|J_{nl}^{q,l+1}|^2}{2l+1} \right),
\end{multline}

where, $E_{nl}$ is energy eigenvalues corresponding to $g_{nl}(r)$, $\alpha_{s}^{u}$ is coupling constant, scaled as $\alpha_{s}^{u} = \alpha_{s}(\alpha_{s}m_{Q}^{2}/2)$ and $m_{Q}$ is the mass of the heavy quark. The $J_{nl}^{ql^{'}}$ is the probability density
obtained by using the singlet $g^*_{nl}(r)$ and octet $h_{ql'}(r)$ wavefunctions as follows, 

\begin{equation}
 J_{nl}^{ql'} = \int_0^\infty r\; g^*_{nl}(r)\;h_{ql'}(r)dr.
\end{equation}

Here, $h_{ql'}(r)$ is the wavefunction of the color octet state of quarkonia. It is obtained by solving the
Schr\"{o}dinger equation with the octet potential $V_{8} = \alpha_{eff}/8r$. The effective coupling constant, $\alpha_{eff}$ is defined at soft scale $\alpha_{s}^{s} = \alpha_{s} (m_{Q} \alpha_{s}/2)$, given as  $\alpha_{eff} = \frac{4}{3}\alpha_{s}^{s}$. The value of $q$ here, is determined using conservation of energy, $q = \sqrt{m_{Q}(E_{g}+E_{nl})}$. Next, we obtained the mean of the gluonic dissociation cross-section to obtain $\Gamma_{gd,nl}$ of a quarkonia state moving with speed $v$. To do so, we thermal averaged over the modified Bose-Einstein distribution function for gluons in the rest frame of quarkonia~\cite{crs2}. Thus, gluonic dissociation is found as,

\begin{equation}
\Gamma_{gd,nl}(\tau,p_{T},b) = \frac{g_d}{4\pi^2} \int_{0}^{\infty}
\int_{0}^{\pi} \frac{dp_g\,d\theta\,\sin\theta\,p_g^2
\sigma_{gd,nl}(E_g)}{e^{ \{\frac{\gamma E_{g}}{T_{eff}}(1 +
v\cos\theta)\}} - 1},
\label{glud}
\end{equation}

where $\gamma$ is a Lorentz factor and $\theta$ is the angle between $v$ and incoming gluon with energy $E_{g}$. $p_T$ is the transverse momentum of the quarkonia and $g_d = 16$ is the number of gluonic degrees of freedom. \\

\subsection{Doppler Shift}
\label{DS}

As a massive particle, the quarkonia may not experience the same temperature as the surrounding medium. The relativistic  Doppler shift induced by the relative velocity ($v_{r}$) between the medium and heavy meson causes an angle-dependent effective temperature ($T_{eff}$), which is expressed as detailed in Refs.~\cite{Escobedo:2013tca, crs2}. As a consequence, the Doppler effect, a blue-shifted $T_{eff}$, is expected in the forward direction and a red-shift in the backward direction. In the blue-shifted region $T_{eff} > T$, however, the blues-shifted is confined to very smaller angles~\cite{Escobedo:2013tca}. While the red-shifted region dominates over the blue-shifted at all relative velocities $v_{r} > 0$; i.e., $T_{eff} < T$~\cite{teff}. As the red-shift grows with increasing velocity, effective temperature decreases further for particles with high  $p_{T}$ values. The $T_{eff}$ is defined as; 

\begin{equation}
T_{eff}(\theta,|v_{r}|) = \frac{T\;\sqrt{1 - |v_{r}|^{2}}}{1 -
|v_{r}|\;\cos \theta},
\label{tt}
\end{equation}

here $\theta$ is the angle between $v_{r}$ and incoming light partons. To
calculate the relative velocity, $v_{r}$, we have taken medium velocity, $v_{m}
= 0.7c$, and quarkonia velocity, $v_{nl} = p_{T}/E_{T}$.
Here $p_{T}$ is transverse momentum of quarkonia and $E_{T} = \sqrt{p_{T}^{2} +
M_{nl}^{2}}$ is its transverse energy, $M_{nl}$ is the mass of corresponding
quarkonium state. We have averaged Eq.(\ref{tt}) over the solid angle and
obtained the average effective temperature given by:

\begin{equation} 
T_{eff}(\tau,b,p_{T}) = T(\tau,b)\;\frac{\sqrt{1 -
|v_{r}|^{2}}}{2\;|v_{r}|}\;\ln\Bigg[\;\frac{1 + |v_{r}|}{1 - |v_{r}|}\Bigg].
\end{equation}

After incorporating the $T_{eff}$ correction, we have obtained the total dissociation decay width by taking the sum over the Eqs. \eqref{cold} and \eqref{glud} correspond to the collisional damping and the gluonic dissociation, respectively.

\begin{equation}
 \Gamma_{D,nl} = \Gamma_{d,nl} + \Gamma_{gd,nl}.
\label{GmD}
\end{equation}
 Next, we shall discuss the regeneration of the bound state in the presence of the anisotropic QGP medium. 

\subsection{Regeneration of quarkonia in the anisotropic QGP medium}
The regeneration via correlated $q-\bar{q}$ pairs is
considered through the de-excitation of the octet state to a singlet state via emitting a gluon. The recombination cross-section $\sigma_{f,nl}$ is obtained from gluonic dissociation cross-section $\sigma_{gd,nl}$ using the detailed balance approach ~\cite{crs1},

\begin{equation}
 \sigma_{f,nl} = \frac{48}{36}\sigma_{gd,nl}
\frac{(s-M_{nl}^{2})^{2}}{s(s-4\;m_{Q}^{2})}.
\end{equation}

Here, $s$ is the Mandelstam variable, $s = ({\bf p_{q}
+ p_{\bar{q}}})^{2}$, where $\bf p_{q}$ and $\bf p_{\bar{q}}$ are four momentum of $q$ and $\bar{q}$, respectively. 
The recombination factor, $\Gamma_{F,nl}$ is obtained by taking the
thermal average of the product of recombination cross-section and relative
velocity $v_{rel}$ between $q$ and $\bar{q}$ as~\cite{thews3}:

\begin{equation}
\Gamma_{F,nl}=<\sigma_{f,nl}\;v_{rel}>_{p_{q}},    
\label{eq:x}
\end{equation}

rewriting Eq.\eqref{eq:x} as,

\begin{equation}
 \Gamma_{F,nl} =
\frac{\int_{p_{q,min}}^{p_{q,max}}\int_{p_{\bar{q},min}}^{p_{\bar{q},max}}
dp_{q}\; dp_{\bar{q}}\; p_{q}^{2}\;p_{\bar{q}}^{2}\;
f_{q}\;f_{\bar{q}}\;\sigma_{f,nl}\;v_{rel}
}{\int_{p_{q,min}}^{p_{q,max}}\int_{p_{\bar{q},min}}^{p_{\bar{q},max}}
dp_{q}\; dp_{\bar{q}}\; p_{q}^{2}\;p_{\bar{q}}^{2}\;
f_{q}\;f_{\bar{q}}},
\label{eq:Fnl}
\end{equation}

where, $p_{q}$ and $p_{\bar{q}}$ are respectively, $3$-momentum of heavy quark and heavy anti-quark
under investigation. The relative velocity of
$q-\bar{q}$ pair in the medium is given as,

\begin{equation}
v_{rel} =
\sqrt{\frac{( {\bf p_{q}^{\mu}\;
p_{\bar{q} \mu}})^{2}-m_{Q}^{4}}{p_{q}^{2}\;p_{\bar{q}}^{2}
+ m_{Q}^{2}(p_{q}^{2} + p_{\bar{q}}^{2}  + m_{Q}^{2})}}.
\end{equation}

The $f_{q,\bar{q}}$ is the modified Fermi-Dirac distribution function of heavy quark and heavy anti-quark given as,
\begin{eqnarray}
f_{q,\bar{q}} =
\lambda_{q,\bar{q}}/(e^{E_{q,\bar{q}}/T_{eff}} + 1).
\end{eqnarray}
Here $E_{q,\bar{q}} = \sqrt{p_{q,\bar{q}}^{2} + m_{Q}^{2}}$ is the energy of heavy quark and
heavy anti-quark, in medium and $\lambda_{q,\bar{q}}$ is their respective
fugacity terms. It should be emphasized that while these fugacity terms do not affect the integration outlined in 
Eq.\eqref{eq:Fnl}. They effectively cancel out from both the numerator and denominator, rendering them unnecessary for the present analysis. Nevertheless, they are included here for the sake of thoroughness and completeness. Next, there is an open discussion on accounting for the Bose Stimulation factor for outgoing gluon in the quarkonia regeneration through the octet-to-singlet transition or gluonic de-excitation. It is argued that the Bose stimulation factor has a minimal impact on the regeneration rate as it merely affects it by less than 10\% ~\cite{Du:2017qkv}. Since the present work investigates quarkonia regeneration within an anisotropic medium, the Bose stimulation factor is omitted in Eq.\eqref{eq:Fnl} to reduce the complexity in the formulation. Next, we shall discuss the results obtained in this analysis.

\section{Results and Discussion}
\label{RaD}

In this study, we investigate how the anisotropic properties of the medium impact the decay width
($\Gamma_{D}$) and recombination reactivity ($\Gamma_{F}$) of $J/\psi$ and $\Upsilon$(1S) with masses 3.1 GeV and 9.46 GeV, respectively. We determine $\Gamma_{D}$ and
$\Gamma_{F}$ using the expressions presented in Eqs.~\eqref{GmD} and ~\eqref{eq:Fnl}, respectively. We 
thoroughly analyze how these quantities depend on the degree of medium anisotropy, $\zeta$. Additionally, we examine 
the relativistic Doppler effect, which arises from particle motion and leads to changes in the effective temperature of 
the particles in relation to the medium. We quantify this effect using the parameter $T_{eff}$ included through 
the color map in all the plots of $\Gamma_D$. Our results demonstrate 
the dependence of $\Gamma_{D}$ on both $T_{eff}$ and transverse momentum ($p_{T}$). We omit the detailed 
discussion of the $T_{eff}$ dependence for $\Gamma_{F}$, as it is difficult to show effective temperature for the 
quark-antiquark pair as depending upon their velocity $q-\bar{q}$ may have different values of $T_{eff}$. However, the 
effective temperature for  $q-\bar{q}$ is considered while calculating the $\Gamma_{F}$. Our analysis considers three 
distinct temperature values, T = 200, 300, and 400 MeV, and two values for anisotropic parameter $\zeta$ = 0.2 and 0.4. 
These specific temperature and anisotropy choices are made to elucidate the qualitative impact of medium anisotropy on 
the dissociation of quarkonia within the QGP medium.

\begin{figure}
\begin{center}
    
\includegraphics[height=6.5cm,width=8.2cm]{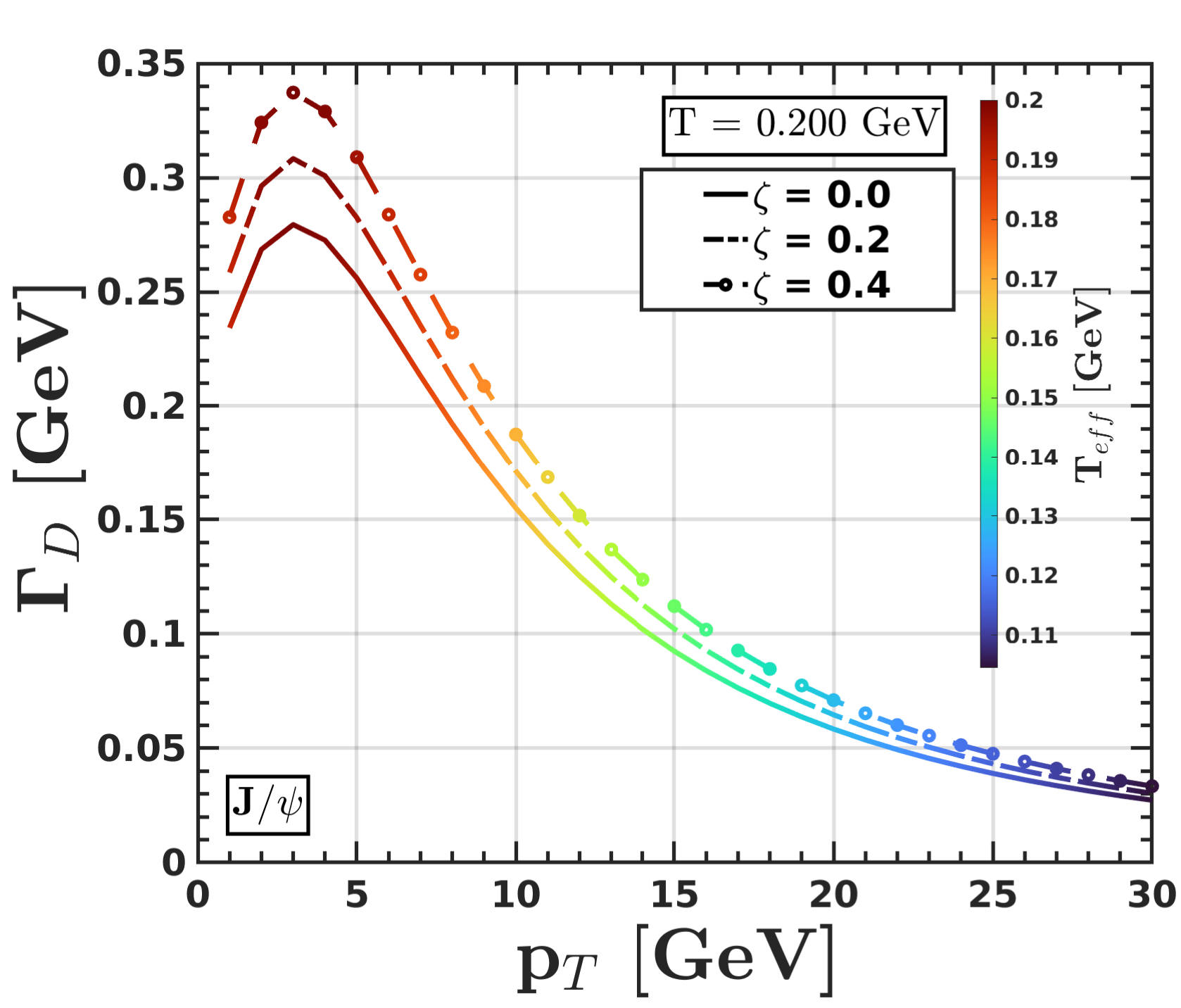}
\includegraphics[height=6.5cm,width=8.2cm]{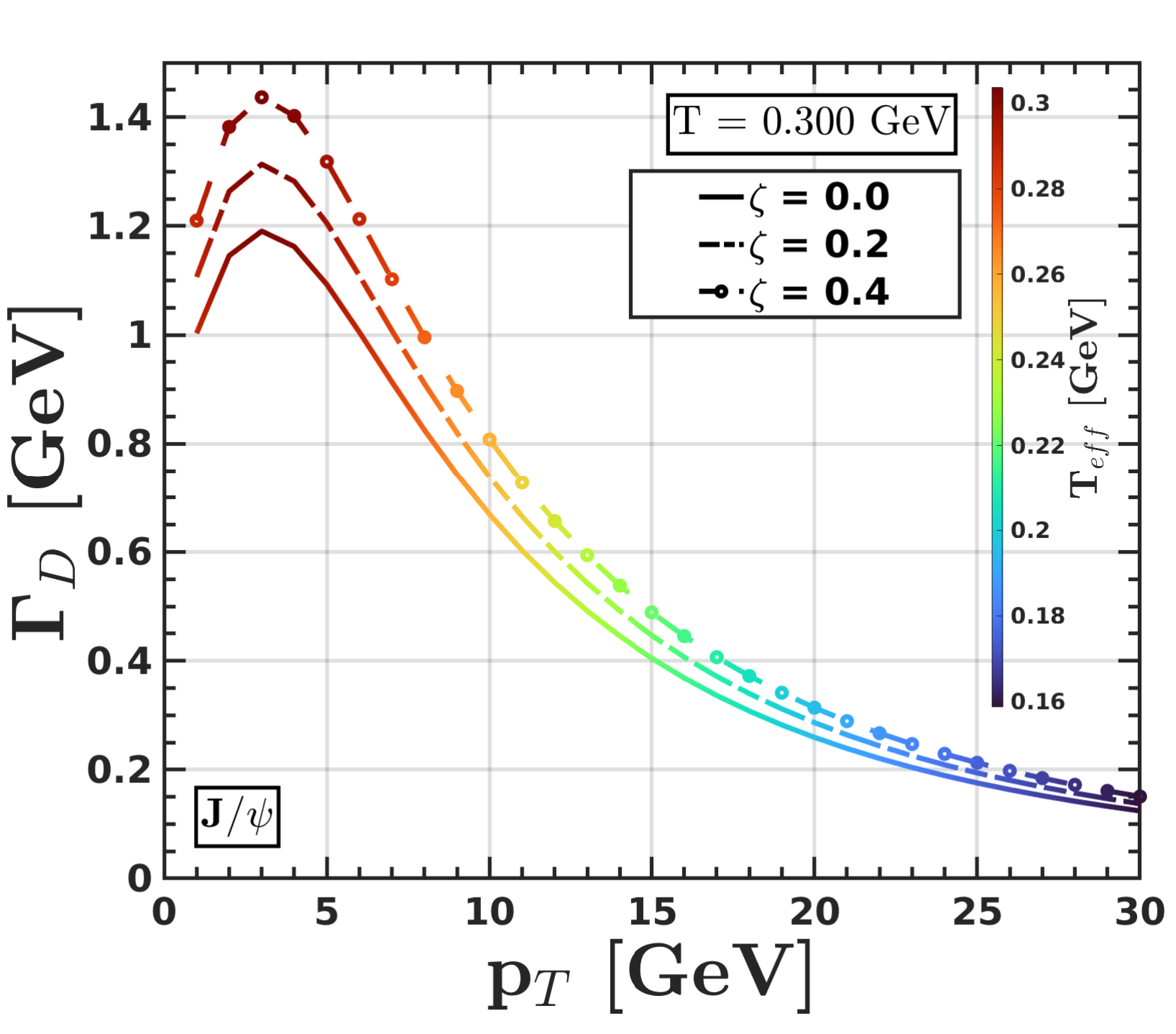}
\includegraphics[height=6.5cm,width=8.2cm]{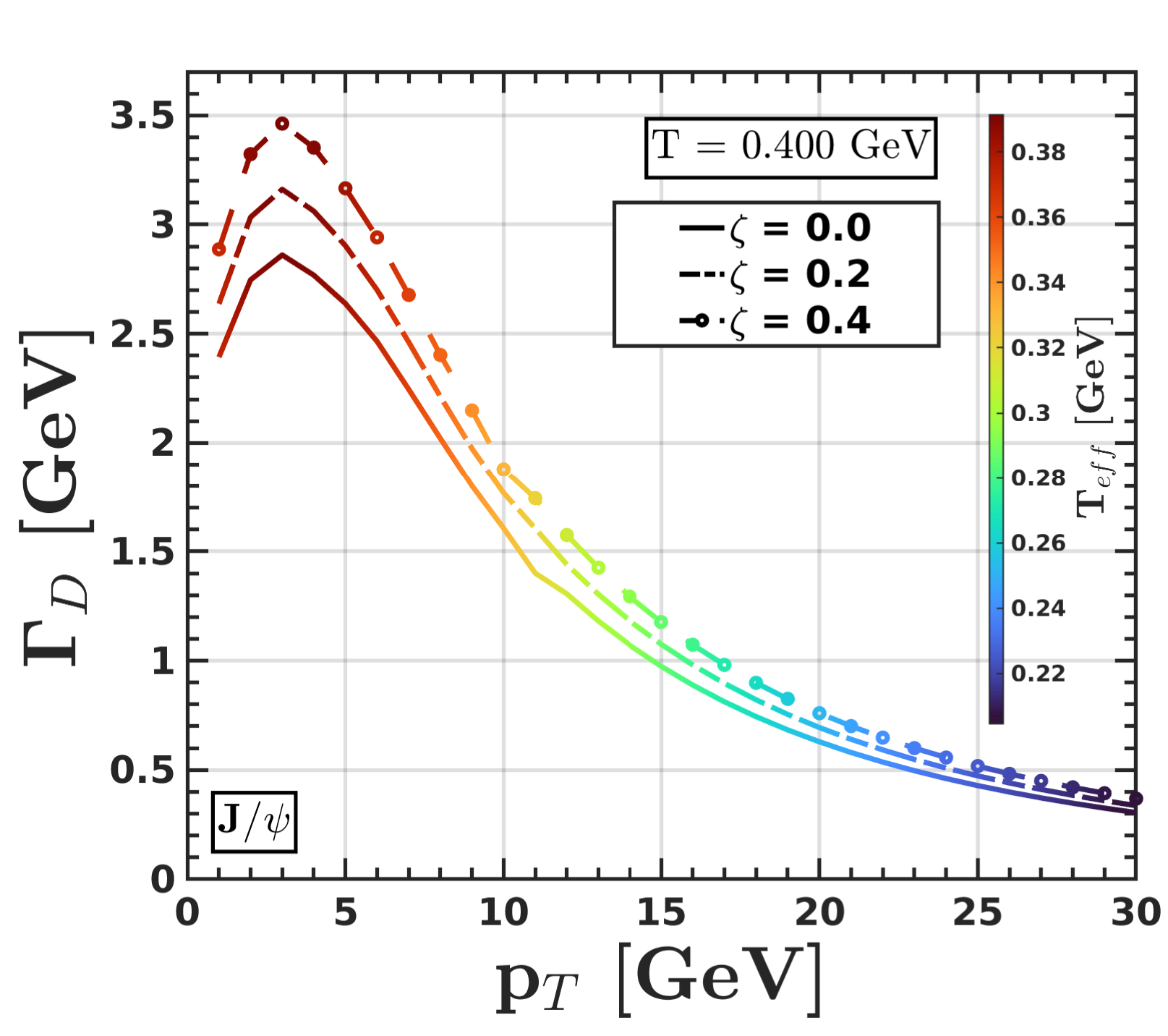}
\caption{(Color Online) The net decay width, $\Gamma_{D}$ for $J/\psi$ as a function of transverse momentum, $p_{T}$ is demonstrated for the anisotropic parameter, $\zeta$ = 0, 0.2, 0.4 and  medium 
temperature, T = 200, 300, 400 MeV. The $\Gamma_D$ with effective temperature, $T_{eff}$ is shown through a color map corresponding to the individual values of T.}
\label{fig:3DGD}
\end{center}
\end{figure} 

Fig.~\ref{fig:3DGD} shows the change  in $\Gamma_D$ for $J/\psi$ with respect to $p_T$ and $T_{eff}$, at different  
values of $T$ and $\zeta$. The  $\Gamma_D$ increases with increasing $\zeta$ from 0 to 0.4  and further grows with 
respect to the change in the medium temperature from 200 MeV to 400 MeV. At low-$p_{T}$ and large $T_{eff}$, the impact 
of anisotropy on $\Gamma_D$ is distinguishable, while with increasing $p_{T}$, this difference diminishes ($T_{eff}$ 
also decreases). Initially, at $p_{T}<3$ GeV, there is a slight increase in the  $\Gamma_D$ for all $\zeta$ values. 
This rising trend is a consequence of the $T_{eff}$, as  $T_{eff}$ increases up to $p_{T}<3$ GeV, and after that, it 
decreases with increasing $p_{T}$, and as a result, $\Gamma_D$ also 
decreases for $p_{T}>3$ GeV. It suggests that the particle dissociation rate will be high if the particle temperature 
is large and comparable to medium temperature. The quarkonia with high-$p_{T}$ quickly traverses through the medium; 
therefore, high-$p_{T}$ $J/\psi$ feels less heat, and survival of such particle becomes more probable.\\

\begin{figure}
\begin{center}
\includegraphics[height=6.5cm,width=8.2cm]{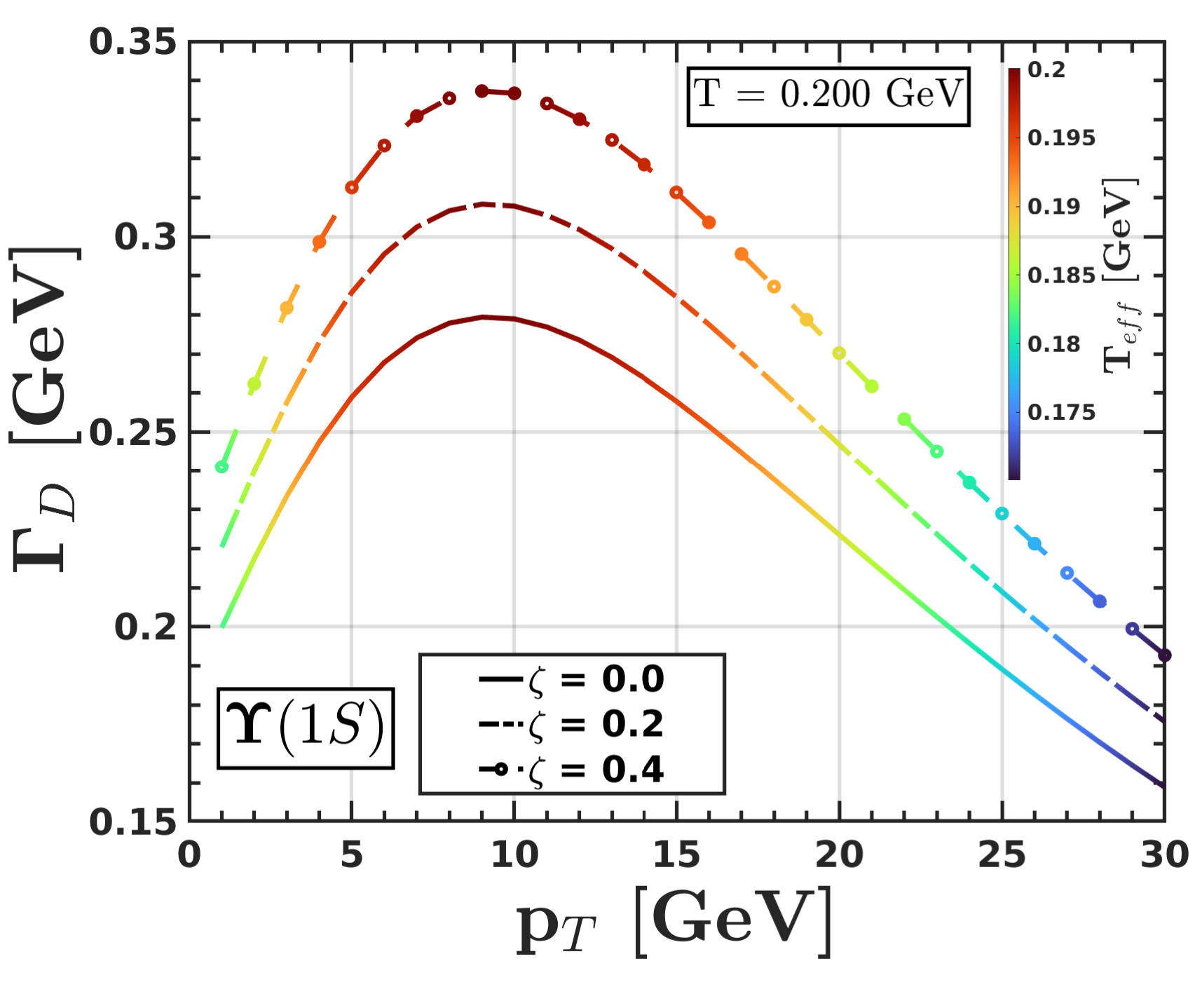}
\includegraphics[height=6.5cm,width=8.2cm]{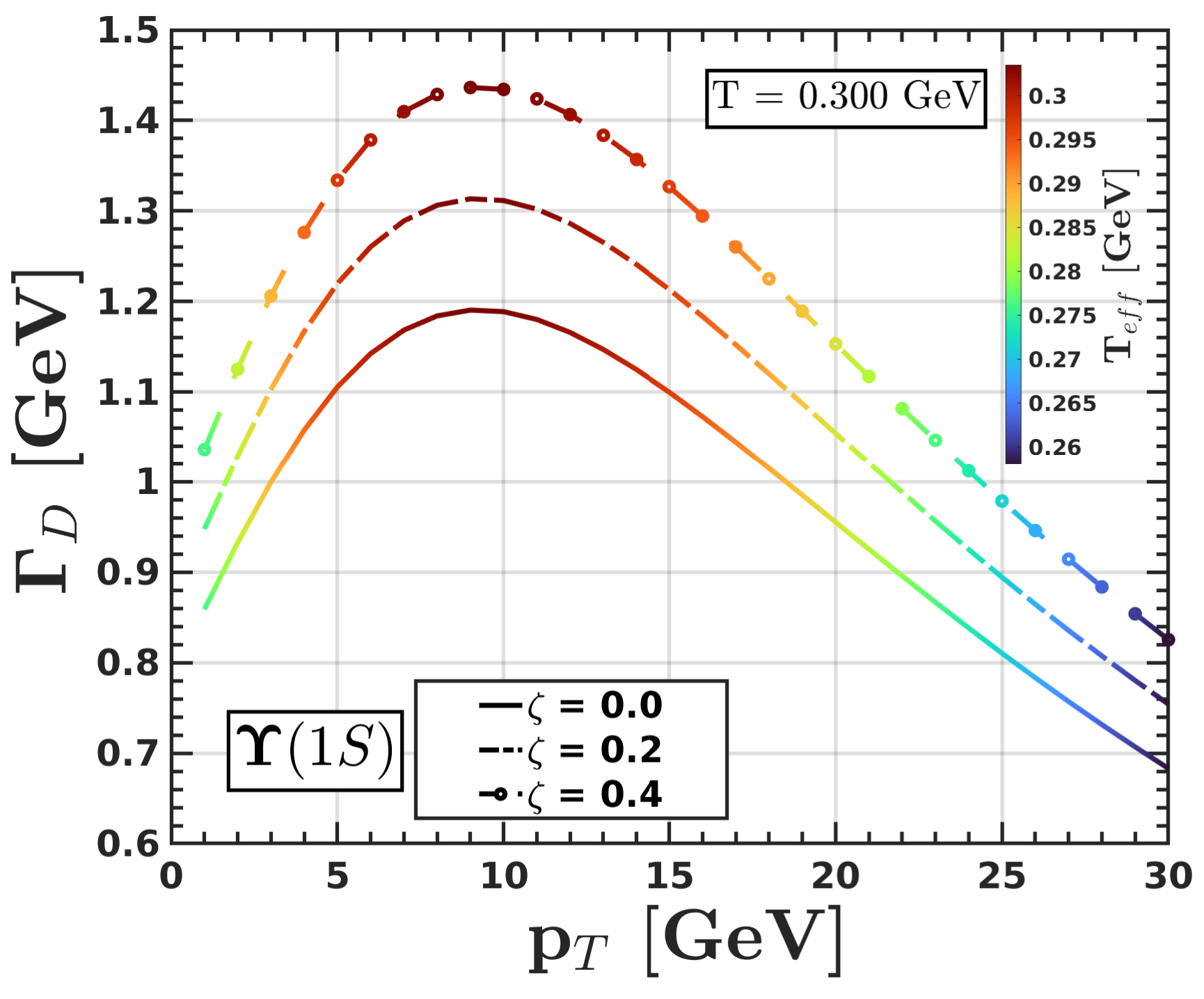}
\includegraphics[height=6.5cm,width=8.2cm]{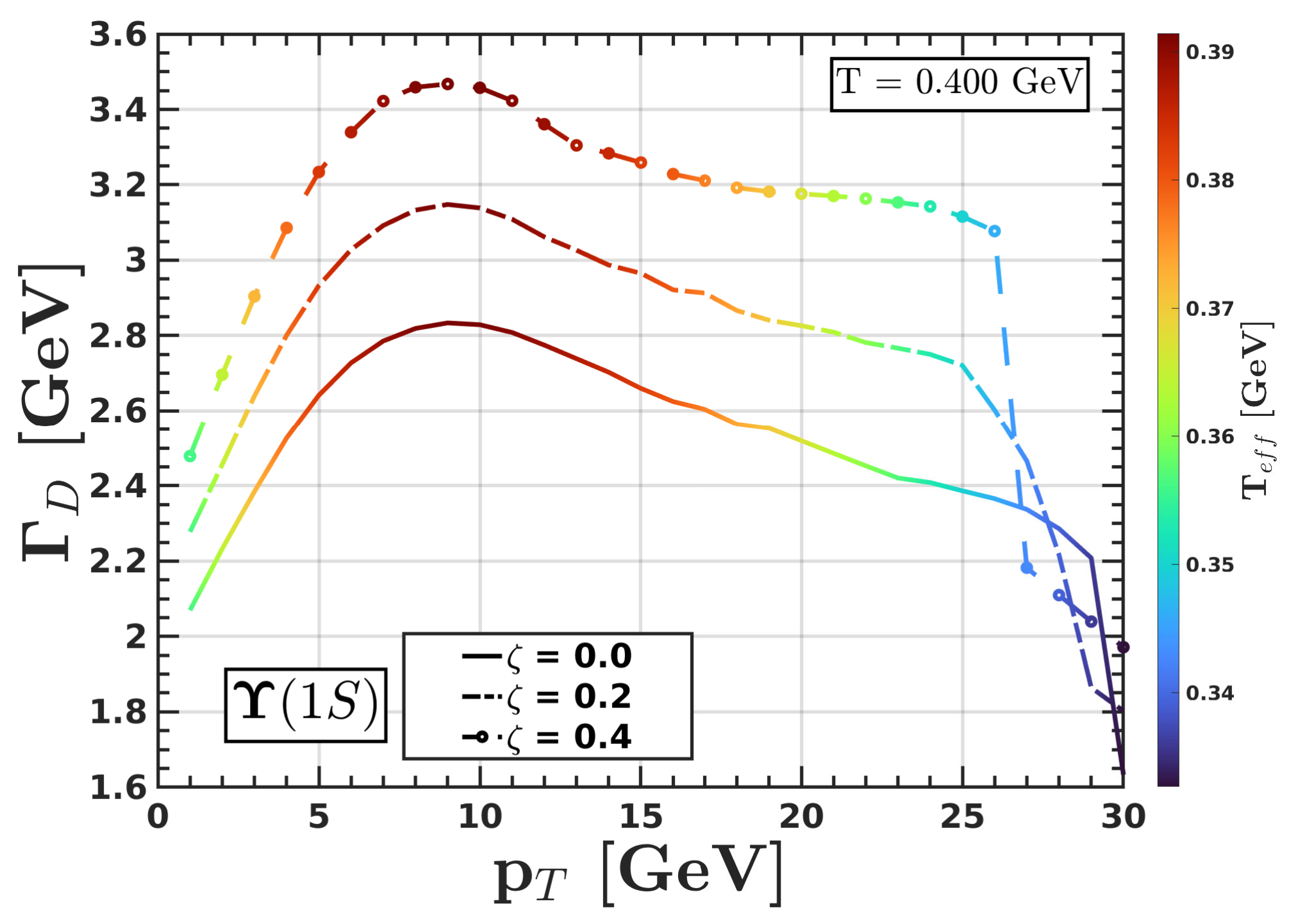}
\end{center}
\caption{(Color Online)  The net decay width, $\Gamma_{D}$ for $\Upsilon$(1S) as a function of transverse momentum, 
$p_{T}$  and effective temperature, $T_{eff}$ is demonstrated for the anisotropic parameter, $\zeta$ = 0, 0.2, 0.4 and medium temperature, T = 200, 300, 400 MeV. The $\Gamma_D$ with effective temperature, $T_{eff}$ is shown through a color map corresponding to the individual values of T.}
\label{fig:3UGD}
\end{figure}

{The decay width of $\Upsilon$(1S) is shown in Fig.~\ref{fig:3UGD}; it follows the same reasoning as  
Fig.~\ref{fig:3DGD} 
except the difference in $\Gamma_D$ corresponding to $\zeta$ values remains consistent throughout $p_{T} = 30$ GeV 
at T = 200 and 300 MeV. While at T = 400 MeV,  $\Gamma_D$ is almost constant up to $T_{eff}\approx$360 
MeV, then a sudden change is observed, further decreasing the $\Gamma_D$. This anomaly in decay width for 
$\Upsilon$(1S) at T = 400 MeV comes because the change in the $T_{eff}$ is almost negligible up to $p_{T}\sim20$ GeV, 
and as a consequence, change in $\Gamma_D$ becomes stagnant at  $p_{T}<20$ GeV. It is to be noted that the difference 
in the analyses of $J/\psi$ and $\Upsilon$(1S) relies on their masses. As a heavy particle, the $\Upsilon$(1S) 
dissociates at higher temperatures than $J/\psi$. Various articles have studied collisional damping, including the anisotropic medium, from different perspectives. However, in most of these analyses, the aspect of gluonic dissociation 
is missing. Fig.~\ref{fig:3DGD} \& \ref{fig:3UGD} contain both the collisional damping and gluonic dissociation 
considering the anisotropy in the QGP medium.}\\

\begin{figure}
\includegraphics[height=6.5cm,width=8.2cm]{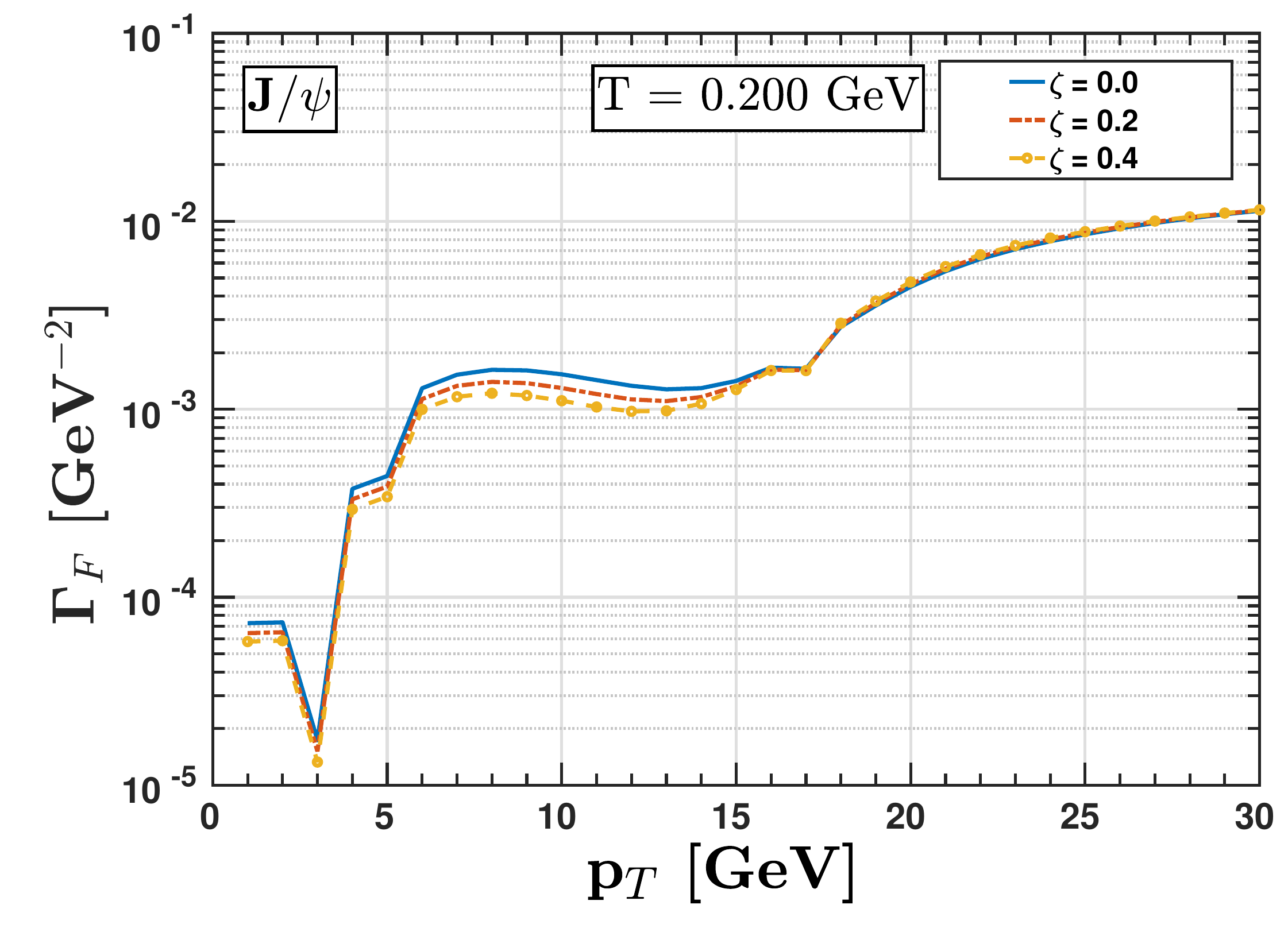}
\includegraphics[height=6.5cm,width=8.2cm]{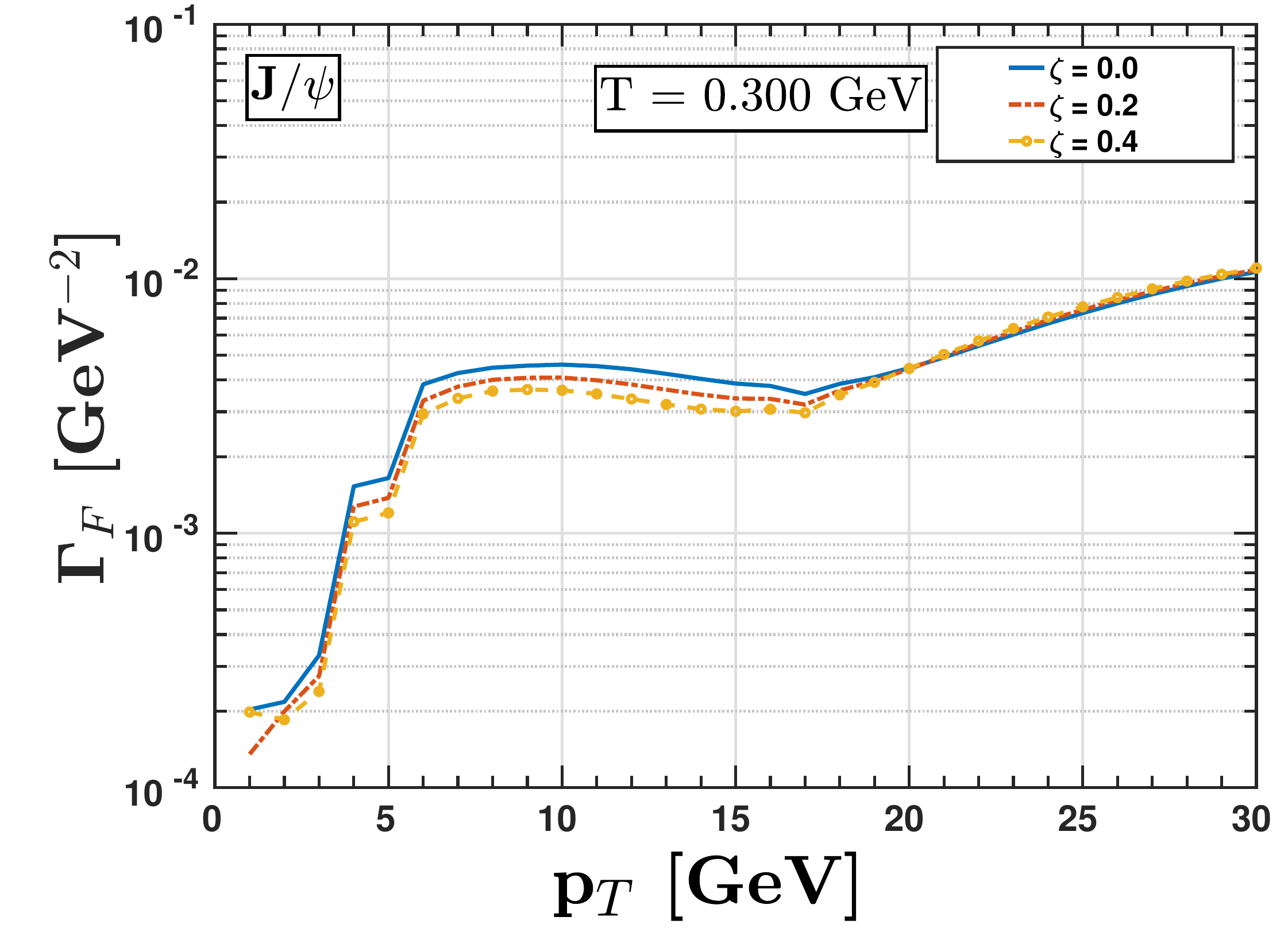}
\includegraphics[height=6.5cm,width=8.2cm]{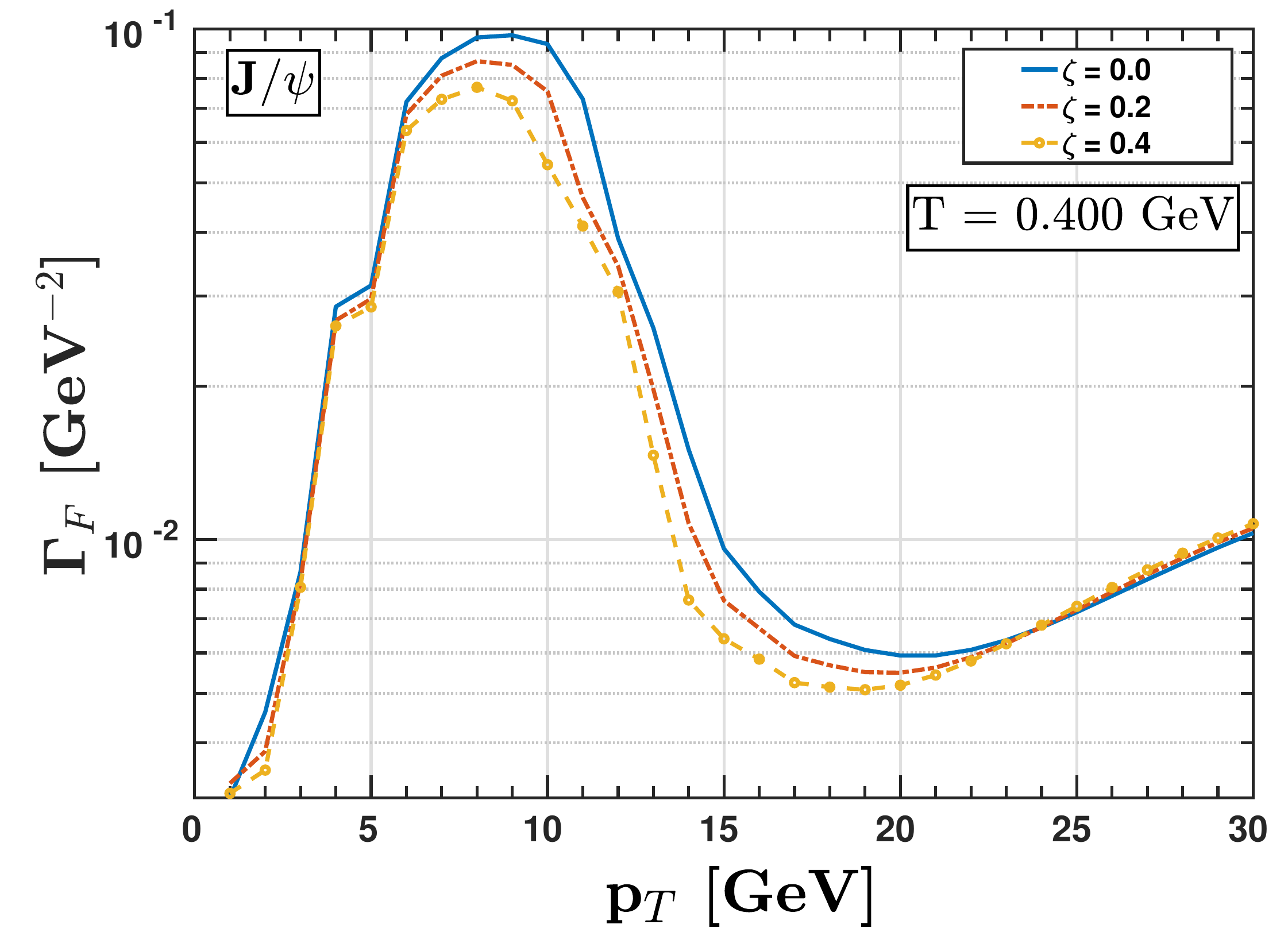}
\caption{(Color Online) The recombination reactivity, $\Gamma_{F}$ for $J/\psi$ is demonstrated in the semilog scale as a function of transverse momentum, $p_T$ with medium anisotropy parameter, $\zeta$ = 0, 0.2, 0.4 for T = 200, 300, 400 MeV.}
\label{fig:JFl}
\end{figure}

In Fig.~\ref{fig:JFl}, the variation of recombination reactivity, $\Gamma_F$ of $J/\Psi$, is shown with respect to $p_T$ 
at different anisotropic strength and temperature values. It is found that $\Gamma_F$ has a non-trivial dependence on  
the three {\it i.e.,} $p_T$, $T$ and $\zeta$. Regeneration of $J/\psi$ is almost independent of medium anisotropy 
at T = 200 MeV. Small dependence on anisotropy can be observed in the intermediate $p_{T}$  ($5<p_{T}<17$ GeV) at T = 
300 
MeV. It depicts that an anisotropic medium is less favorable to $J/\psi$ regeneration than an isotropic medium. As 
anisotropy induces or supports $J/\psi$ dissociation, it is obvious that it would restrict the particle regeneration, 
though its impact is negligible. Regeneration increases with increasing $p_{T}$, except at mid-$p_{T}$ where 
particle interactions and large medium temperature constrain the $J/\psi$ to form. While at $p_{T}>17$ GeV  $c-\bar{c}$ 
pair feels a lower temperature than the medium, and as they move at a high speed, they easily traverse 
through the medium. Consequently, we get a linear increase in regeneration reactivity with $p_{T}$. As we move to a 
very 
high temperature, T= 400 MeV, the probability of recombination suppresses to a very narrow range of $p_T$. Here, 
$\Gamma_F$ showed a non-monotonic behavior, as medium temperature T = 400 MeV is larger than the dissociation 
temperature of $J/\psi$ ($T_{D}\sim$350 MeV); therefore a large number of $c-\bar{c}$ pair are expected, and as the 
interaction probability is large at mid-$p_{T}$, most of the $c-\bar{c}$ recombine to form $J/\psi$. Hence, a maximum 
recombination probability is attained at $p_{T}\sim8-10$ GeV for the isotropic case. At $p_{T}>15$ GeV for T = 400 MeV, 
the order of recombination becomes almost equivalent to the T = 200 and 300 MeV, following a similar explanation as the 
previous one corresponding to high-$p_{T}$. Moreover, regeneration due to un-correlated $q-\bar{q}$ pair dominates at 
low-$p_{T}$ and decreases rapidly at high-$p_{T}$. Thus, charmonia regeneration due to correlated $c-\bar{c}$ pair is 
contrary to the regeneration due to un-correlated $q-\bar{q}$ pair.\\

\begin{figure}
\includegraphics[height=6.5cm,width=8.2cm]{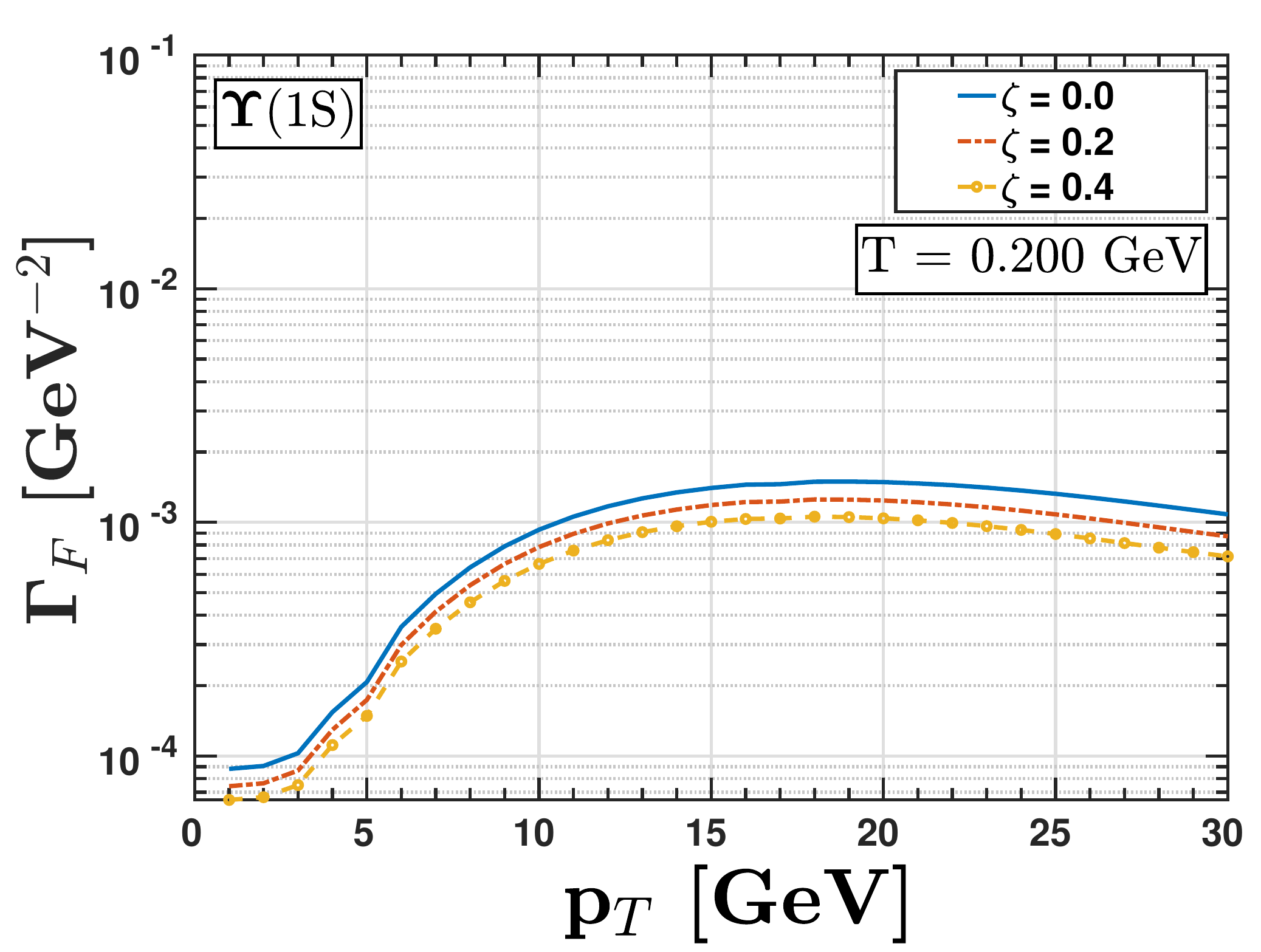}
\includegraphics[height=6.5cm,width=8.2cm]{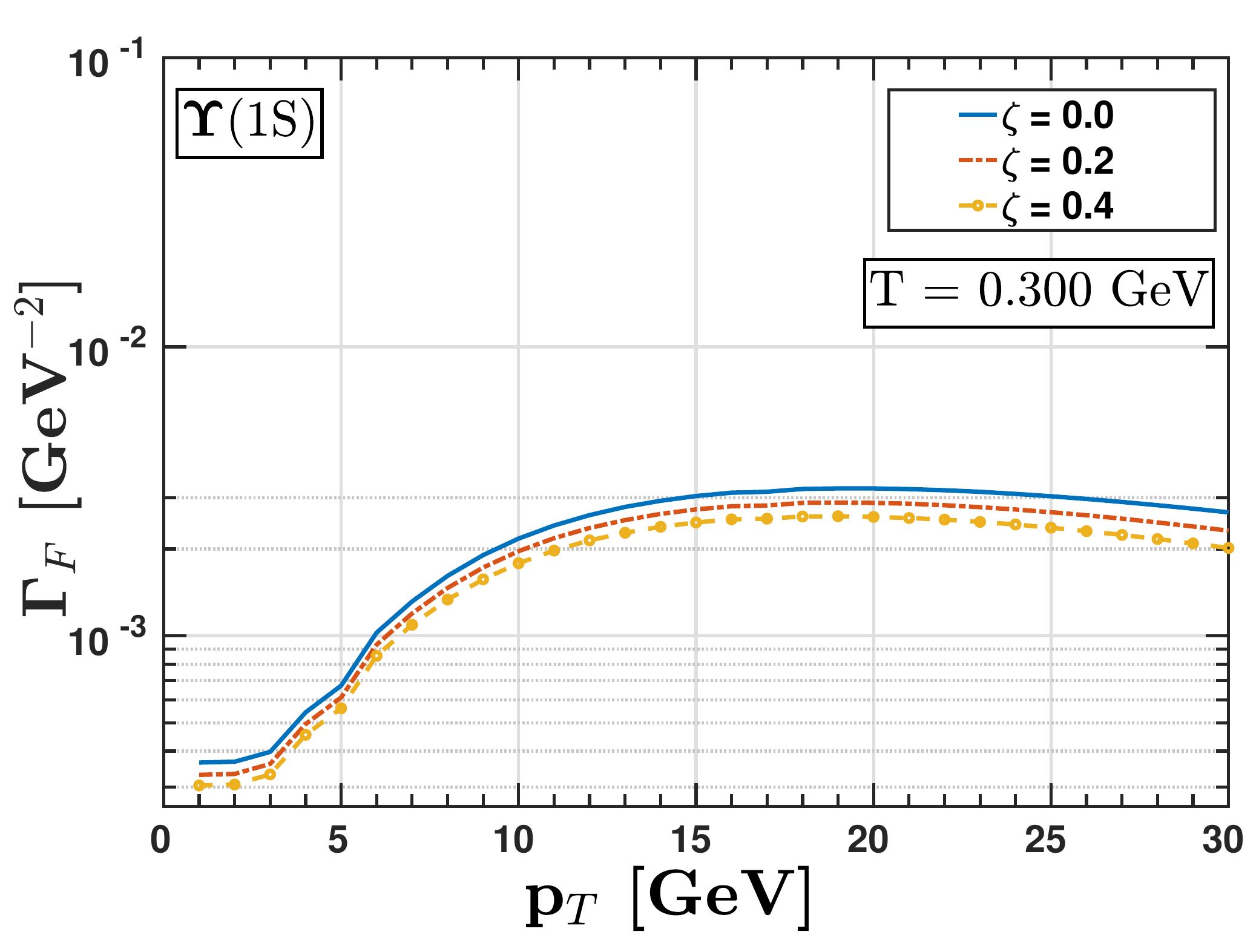}
\includegraphics[height=6.5cm,width=8.2cm]{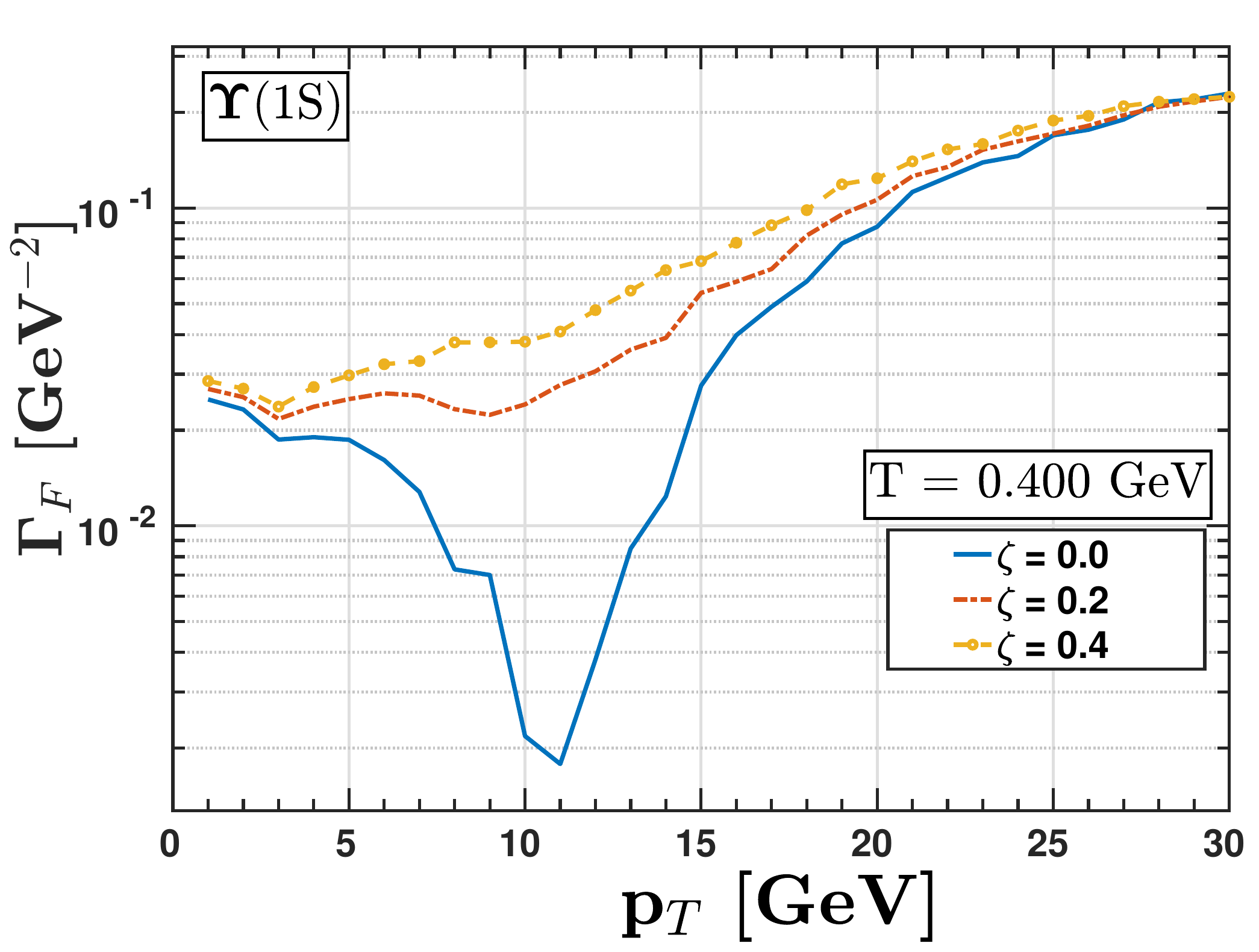}
\caption{(Color Online) The recombination reactivity, $\Gamma_{F}$ for $\Upsilon$(1S) is demonstrated  in the semilog scale  as a function of transverse momentum, $p_T$  with medium anisotropy parameter, $\zeta$ = 0, 0.2, 0.4 for T = 200, 300, 400 MeV.}
\label{fig:UFl}
\end{figure}


The regeneration of $\Upsilon$(1S) is depicted in Fig.~\ref{fig:UFl}, there is a slight increase in $\Gamma_F$ 
magnitude 
at T = 300 MeV than 200 MeV temperature. However, it follows the same trend at both temperature values. The effect 
of 
anisotropy on $\Upsilon$(1S) regeneration is visible, though it decreases with increasing temperature. As shown in the 
figure, at T = 200 and 300 MeV, $\Upsilon$(1S) regeneration is dominated for the isotropic medium, which is equivalent 
to $J/\psi$ as shown in  Fig.~\ref{fig:JFl}. Initially, $\Gamma_F$ increases because of the abundance of $b-\bar{b}$ 
octet states at mid-$p_{T}$ range but at high-$p_{T}$, this abundance decreases for low temperatures, and so does the 
$\Gamma_F$. The physics of $\Upsilon$(1S) regeneration probability at T = 400 MeV is as follows: since gluonic 
dissociation 
increases with the increase in temperature, it produces a significant number of $b-\bar{b}$ octet states for 
temperature 
more than $300$ MeV. Such that the de-excitation of $b-\bar{b}$ octet states to $\Upsilon(1S)$ enhances the 
regeneration 
of $\Upsilon(1S)$ at T = 400 MeV, which can be seen in Fig~\ref{fig:UFl}, where the value of $\Gamma_{F}$ is higher at 
T 
= 400 MeV as compared with T = 300 MeV. It also shows that high temperature produces enough  $b-\bar{b}$ pair even at 
high-$p_{T}$ and therefore, we get an increasing pattern in $\Upsilon$(1S) regeneration with $p_{T}$. As gluonic 
excitation decreases at high-$p_{T}$, so the de-excitation of $b-\bar{b}$ octet state into $\Upsilon(1S)$ becomes more 
probable with increasing $p_{T}$ at T = 400. For isotropic case ($\zeta = 0$), $\Gamma_{F}$ decreases between $p_{T}$ 1 
to 10 GeV, because as it is shown in Fig.~\ref{fig:3UGD} the dissociation $\Upsilon$(1S) is more favorable at this 
$p_{T}$ range. Meanwhile, $\Gamma_{F}$ for $\Upsilon$(1S) always has a rising trend for the anisotropic case at T = 400 
MeV. It seems that at high temperatures where anisotropy increases the decay width but also favors the $\Upsilon$(1S) to 
regenerate. Due to high dissociation temperature, $\Upsilon$(1S) can regenerate even at T = 400 MeV or more, and 
therefore, $\Gamma_{F}$ for $\Upsilon$(1S) is almost two times larger than $J/\psi$.

\section{Summary}

This research investigates the modification in quarkonia potential induced by momentum anisotropy within the anisotropic 
hot QCD medium. This anisotropic medium intricately influences the gluon distribution, leading to significant changes 
in quarkonia dissociation and regeneration phenomena. We comprehensively examine the interplay of dissociation 
mechanisms, including gluonic dissociation and collisional damping, considering various strengths of the medium anisotropy through the  
parameter, $\zeta$. Moreover, we meticulously explore the role of the relativistic Doppler effect (RDE) on quarkonia 
dissociation by considering the effective temperature, $T_{eff}$.

Our findings reveal distinct effects of medium anisotropy and RDE on the dissociation and regeneration of $J/\psi$ and 
$\Upsilon$(1S). Notably, $\Upsilon$(1S) dissociation is enhanced in the presence of anisotropy at all temperature 
levels, while $J/\psi$ is more profoundly affected at higher temperatures and lower transverse momentum ($p_{T}$) 
values. The influence of anisotropy on the regeneration of $J/\psi$ is marginal, whereas $\Upsilon$(1S) regeneration is 
particularly dependent on the degree of medium anisotropy. We have observed that medium anisotropy increases and 
supports particle dissociation by reducing its binding strength within the QGP medium as it alters the particle's 
potential. Therefore, if the medium is anisotropic, then quarkonia states become more vulnerable and can have a higher 
decay width than an isotropic medium.\\

The primary objective of this study was to elucidate the effects of medium anisotropy at the recombination process
at various temperatures. These insights can serve as corrections for future investigations into the total survival probability of bottomonium and charmonium in heavy-ion collisions at facilities such as RHIC at Brookhaven National Laboratory (BNL) and LHC at the European Organization for Nuclear 
Research (CERN). As the future upgrades at the LHC focus on the measurement of heavy flavors (along with some others) with increased 
low-$p_{T}$ reach and vertexing efficiencies close to the interaction point, this study will help the theoretical model tuning to better understand quarkonia production/decay dynamics at the RHIC and LHC energies.

Lastly, an outstanding article investigated the regeneration of bottomonia through an open quantum systems approach \cite{Brambilla:2023hkw}, and we duly recognize the sophistication of this method. However, our current approach diverges slightly, and we have yet to derive the survival probability, as it demands additional models beyond our present scope. Consequently, it proves challenging to furnish a direct comparison with the open system approach. Thus, we defer this aspect to future extensions of our work, enabling a thorough comparison with the open system approach.

\label{SaF}

\section{Acknowledgement}
Captain R. Singh and Raghunath Sahoo gratefully acknowledge the DAE-DST, Government of India funding under the mega-science project ``Indian Participation in the ALICE experiment at CERN" bearing Project No. SR/MF/PS-02/2021-IITI (E-37123). MY Jamal would like to acknowledge the SERB-NPDF (National postdoctoral fellow) File No. PDF/2022/001551.

\end{document}